\documentclass[rnote]{aa}

\usepackage{times}
\usepackage{graphicx}
\usepackage{xspace}
\usepackage{epsfig}
\usepackage{natbib}
\usepackage{rotating}
\usepackage{dcolumn}

\newcommand{\gequ}{$\gamma$\,Equ\xspace}

\begin{document}

\title{Search of X-ray emission from roAp stars: The case of $\gamma$\,Equulei}

%\subtitle{blablabla} 

\author{B. Stelzer \inst{1} \and C.A. Hummel \inst{2} \and M. Sch\"oller \inst{2} \and S. Hubrig \inst{3} \and C. Cowley \inst{4}}

\offprints{B. Stelzer}

\institute{INAF - Osservatorio Astronomico di Palermo,
% INST 1
  Piazza del Parlamento 1,
  I-90134 Palermo, Italy \\ \email{B. Stelzer, stelzer@astropa.inaf.it} \and
% INST 2
  European Southern Observatory, Karl-Schwarzschild-Strasse 2, D-85748 Garching, Germany \and
% INST 3
  Astrophysikalisches Institut Potsdam, An der Sternwarte 16, D-14482 Potsdam, Germany \and
% INST 4
  Department of Astronomy, University of Michigan, Ann Arbor, MI 48109-1042, U.S.A.
}

%\titlerunning{X-ray emission from Z\,CMa during outburst and from its jet}

\date{Received $<$06-12-2010$>$ / Accepted $<$24-02-2011$>$}

\abstract{Rapidly oscillating Ap (roAp) stars represent a subclass of magnetic, 
chemically peculiar stars. The explanation for their pulsations includes
suppressed convection due to the strong magnetic field. These
stars rotate slowly such that a solar-like dynamo and ensuing magnetic
activity is unlikely to be present. 
On the other hand, magnetic activity could provide the particle 
acceleration suspected to be responsible for the presence of short-lived 
radionuclides on some roAp stars. 
} 
% context heading (optional)
{The detection of X-ray emission from Ap stars can be an indicator 
for the presence of magnetic activity and dynamo action, 
provided different origins for the emission, such as wind shocks and 
close late-type companions, can be excluded. 
Here we report on results for \gequ, 
the only roAp star for which an X-ray detection is reported in ROSAT catalogs. 
}
% aims heading (mandatory)
{We use high resolution imaging in X-rays with Chandra and in the near-infrared with 
NACO/VLT that allow us to spatially resolve companions down to 
$\leq 1^{\prime\prime}$ and $\sim 0.06^{\prime\prime}$ separations, respectively. 
}
% methods heading (mandatory)
{
%The X-ray image presents possible indications for faint X-ray emission from the 
The bulk of the X-ray emission is associated with a companion of \gequ 
identified in our NACO image.  
Assuming coevality with the primary roAp star ($\sim 900$\,Myr), 
the available photometry for the companion points at a
K-type star with $\sim 0.6\,M_\odot$. Its X-ray properties 
are in agreement with the predictions for its age and mass. 
An excess of photons 
with respect to the expected background and contribution from the nearby 
companion is observed near the optical position of \gequ. 
We estimate an X-ray luminosity of $\log{L_{\rm x}}\,{\rm [erg/s]} = 26.6$ and
 $\log{(L_{\rm x}/L_{\rm bol})} = -7.9$ for this emission. 
A small offset 
between the optical and the X-ray image
leaves some doubt on its association with the roAp star. 
}
% results heading (mandatory)
{
The faint X-ray emission that we tentatively ascribe to the roAp star 
is difficult to explain as
a solar-like stellar corona due to its very low $L_{\rm x}/L_{\rm bol}$ level and the very
long rotation period of \gequ. 
It could be produced in magnetically confined wind shocks implying
a mass loss rate of $\sim 10^{-14}\,M_\odot{\rm /yr}$ 
or from an additional unknown late-type companion at separation 
$\leq 0.4^{\prime\prime}$.
If confirmed by future deeper X-ray observations
this emission could point at the origin for the presence of radioactive elements
on some roAp stars. 
}
% conclusions heading (optional)
{
}

\keywords{X-rays: stars -- stars: chemically peculiar -- stars: activity -- stars: binaries: close -- stars: individual: \gequ}

\maketitle

\section{Introduction}\label{sect:intro}

Solar-like stellar magnetic activity relies on rotation and 
the presence of a convection zone.
According to interior models based on mixing-length theory the transition from
radiative to convective
envelope takes place near $T_{\rm eff} \sim 8300$\,K 
\citep[e.g.][]{ChristensenDalsgaard00.1}. 
This is in rough accordance with X-ray and UV  
observations, which place the onset of significant chromospheric and coronal emission somewhere between 
spectral type A7 and F4 \citep{Schmitt85.1, Simon94.1, Schmitt97.1}. 
However, 
as a consequence of the low spatial resolution of the early-day X-ray instruments, 
the contaminating X-ray emission from a possible (or known) low-mass binary companion 
has always been a problem in confirming intermediate-mass stars as intrinsic X-ray 
sources \citep[e.g.][]{Drake94.1}. 

Up to 20\,\% of A and B main-sequence (MS) stars are `chemically peculiar' (CP) stars.
These stars show enhanced lines of some elements in their spectra which are ascribed to 
atmospheric abundance anomalies. 
A sub-group of the CP stars possess strong predominantly dipolar magnetic 
fields \citep{Borra82.1}. 
It is unclear if these fields are fossil relics of the star formation 
process or if they have a dynamo origin. 

The rapidly oscillating Ap stars \citep[roAp stars;][]{Kurtz82.1} 
hold special promise for elucidating the potential of dynamos and ensuing  
magnetic activity 
within the CP class: 
The roAp stars have a photometrically and spectroscopically
estimated temperature range from 8400\,K down to 6800\,K corresponding to  
spectral types A5 to F2.
A shallow convective layer is expected to be present. 
However, 
these stars have large-scale organized magnetic fields, and suppression 
of convection by the magnetic field has been invoked to explain the mechanism 
of the excitation of their pulsations \citep{Balmforth01.1}. 
Consistent with the expected weakness of convection, 
no evidence for chromospheres have been found so far 
\citep{Shore87.1, Seggewiss90.1}. 
%Moreover, 
On the other hand, the possible detection of Pm\,II lines with short half-life 
may be explained by flaring activity \citep{Fivet07.1}.  
None of the roAp stars is a spectroscopic binary \citep{Hubrig00.2}
facilitating the interpretation of eventually detected signs of activity. 

X-ray emission is one of the most efficient indicators of magnetic activity. 
In this paper we present the results from 
a {\em Chandra} observation of the roAp star \gequ,
the slowest rotator of the CP class and, thus, the least likely to exhibit
activity generated by a rotation-dependent dynamo. 
We introduce the target in Sect.~\ref{sect:target}.
In Sects.~\ref{sect:companion}
and~\ref{sect:orbit} we provide the findings from our investigation of 
the nature of our target.
The X-ray observations and their results are described in 
Sect.~\ref{sect:xraydata}. 
A synthesis and our conclusions are given in Sect.~\ref{sect:discussion}.

\section{The target, \gequ}\label{sect:target}

What made us suspicious of magnetic activity on \gequ is 
the detection of a nearby X-ray source 
the {\em ROSAT} Faint Star Catalogue (FSC).  
The offset between the FSC X-ray position and the optical position of 
\gequ is $28.1^{\prime\prime}$, 
i.e. at the limit of the {\em ROSAT} error box. 
\gequ shows no signs for radial velocity variability due to orbital motion, 
and thus it has no spectroscopic companions \citep{Hubrig00.2}. 
However, it is known to be a close visual binary 
(see Sect.~\ref{sect:companion})
that can be resolved in X-rays only with {\em Chandra}. 

Measurements of the longitudinal magnetic field of \gequ 
over nearly six decades show
a clear sinusoidal pattern for which \cite{Bychkov06.1} derived 
a period of $91.3 \pm 3.6$\,yrs. 
Rotational modulation of an inclined dipole is the most widely favored 
scenario for explaining this presumably cyclic variability, 
implying extremely slow rotation for \gequ, and indeed making it the
slowest rotator among all magnetic CP stars. 

For our study we adopt the 
fundamental parameters of \gequ from \cite{Hubrig07.1}:   
$\log{T_{\rm eff}}\,{\rm [K]} = 3.882 \pm 0.017$ 
and $\log{(L/L_\odot)} = 1.144 \pm 0.049$ with the Hipparcos parallax 
$28.38 \pm 0.90$\,mas \citep{Perryman97.1}. 
We note that the stellar parameters derived by \cite{Perraut11.1} 
combining interferometric
measurements of the star's angular diameter with modeling of the spectral 
energy distribution 
are compatible within the error bars with those given by \cite{Hubrig07.1}. 
\cite{Hubrig07.1} have estimated the mass 
and age 
of \gequ
interpolating the solar-abundance \cite{Schaller92.1} tracks. 
In this paper we resort to the MS evolutionary calculations of 
\cite{Pietrinferni04.1}. These models have the advantage of covering 
a wide range of masses, $0.5...10\,M_\odot$, such that they can be applied to 
both \gequ and its fainter visual companion (see Sect.~\ref{sect:companion}). 
Isochrones and tracks of these models can be downloaded from the BaSTI database\footnote{The {\em Bag of Stellar Tracks and Isochrones} database is accessible at http://albione.oa-teramo.inaf.it/} for various sets of parameters.
We use the standard solar model with initial He abundance of $Y=0.273$
and metallicity $Z=0.0198$. For stars with mass above $1.1\,M_\odot$ 
overshooting from the convective core is taken into account in these calculations. 
Interpolating these models we find for \gequ a mass of $1.81 \pm 0.04\,M_\odot$
and an age of $0.88 \pm 0.17$\,Gyr. The uncertainties on the mass
and age are derived considering the uncertainties of $T_{\rm eff}$ and
$L_{\rm bol}$ given by \cite{Hubrig07.1}. Our results are consistent with
the mass and age obtained by \cite{Hubrig07.1}.

\section{The nature of the companion}\label{sect:companion}

\begin{figure}[t]
\begin{center}
\parbox{9.0cm}{
\parbox{4.7cm}{
\resizebox{4.7cm}{!}{\includegraphics{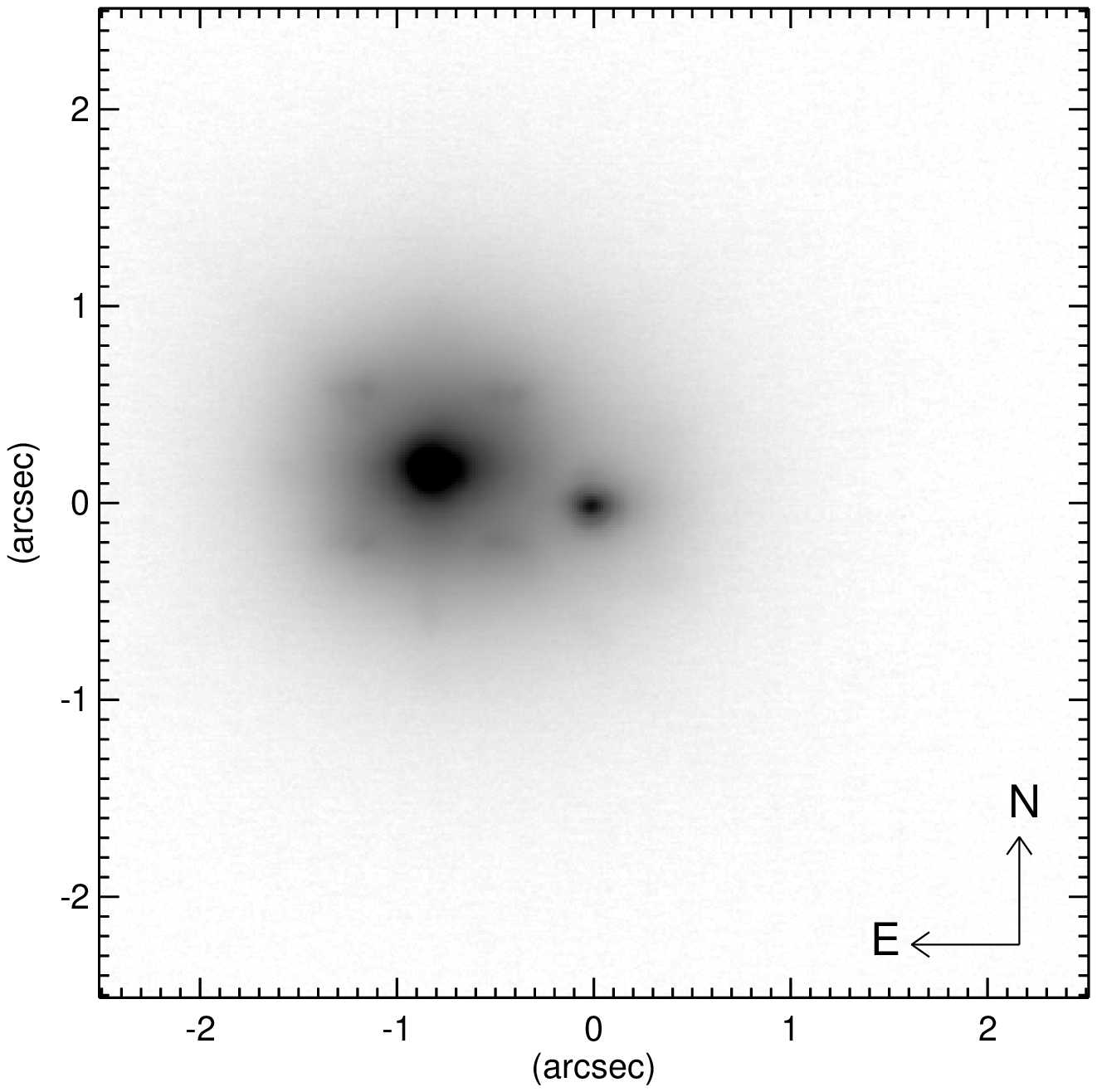}}
}
\parbox{4.3cm}{
\resizebox{4.3cm}{!}{\includegraphics{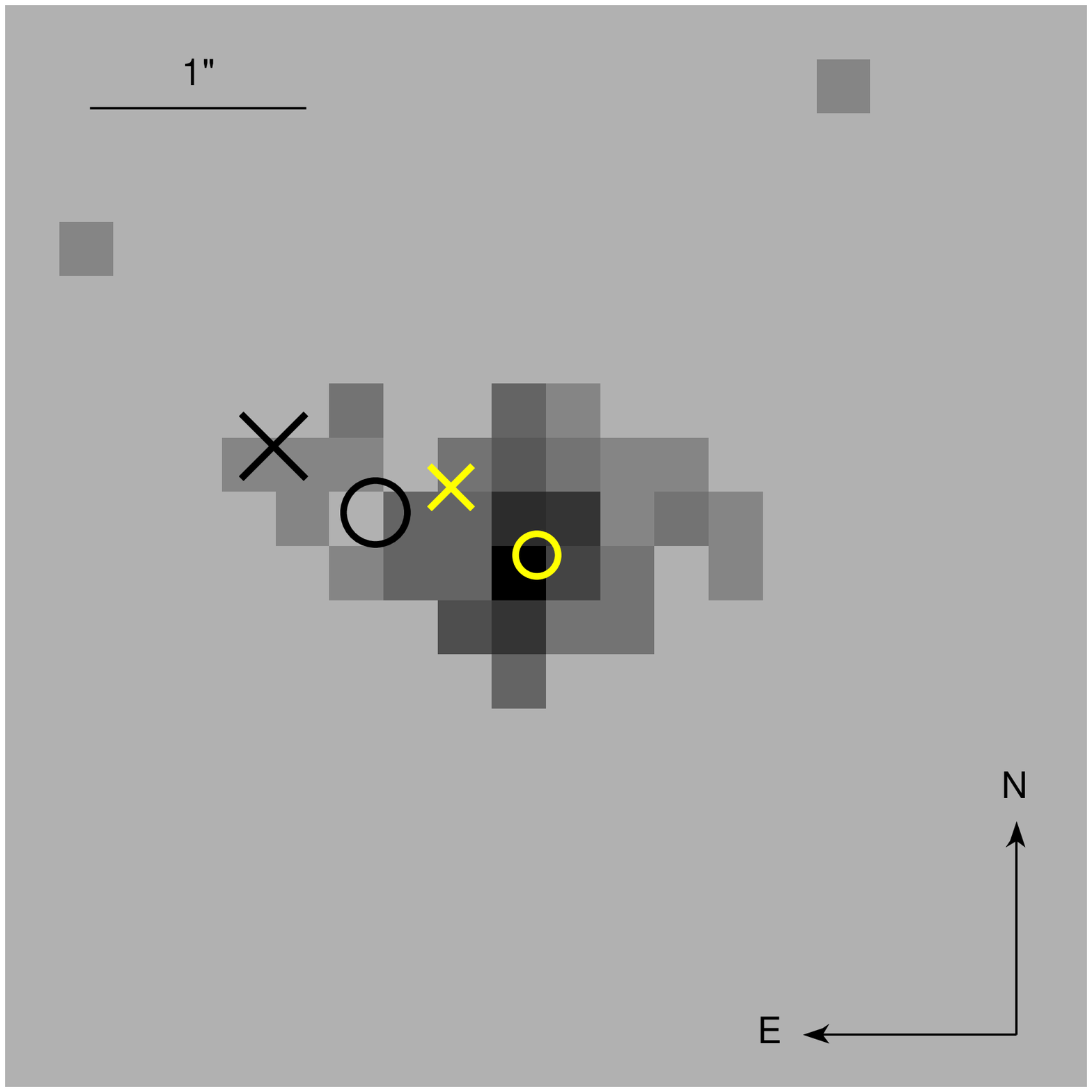}}
}
}
\caption{
[left] NACO $K$-band image centered on \gequ\,B. 
[right] X-ray image in $0.3-8.0$\,keV with $0.25^{\prime\prime}$ pixel size. 
The position of the X-ray source is highlighted with a yellow (small) circle. Positions indicated by x-shaped symbols denote the proper motion corrected {\em Hipparcos} position of \gequ~A at the time of the {\em Chandra} observation (black) and the corresponding expected position of the companion according to the orbit solution (yellow and smaller). 
The expected position of the primary, \gequ~A, 
as inferred from the orbit solution for the assumption that the companion is located at 
the X-ray position, is shown as black (larger) circle. 
}
\label{fig:images}
\end{center}
\end{figure}

The Washington Double Star Catalogue \citep[henceforth WDS; ][]{Mason01.2} 
lists three companions for \gequ. 
The closest companion was also detected in recent adaptive optics imaging 
with NACO at the VLT (Sch\"oller et al., in prep.)
at a separation of $a = 0.826^{\prime\prime} \pm 0.003^{\prime\prime}$ 
and position angle of $\theta = 256.8^\circ \pm 0.5^\circ$; 
see Fig.~\ref{fig:images} (left). 
Sch\"oller et al. estimate  $K=4.09 \pm 0.26$\,mag for the primary 
and $K=6.80 \pm 0.29$\,mag for the secondary.

To constrain the nature of the companion we searched for further information
in the literature. We extracted optical photometry from the Tycho
catalog \citep{Tycho}:
$B_{\rm T} = 9.85 \pm 0.03$\,mag and $V_{\rm T} = 8.69 \pm 0.02$. 
These values can be converted to the Johnson system using the transformations
given in the Introduction to the Hipparcos and Tycho Catalogues. 
We assume coevality with \gequ\,A,
i.e. an age of $\sim 0.88$\,Gyr, in order to derive the stellar parameters
from the evolutionary models. \cite{Pietrinferni04.1} provide $B-V$ colors
for each grid point in their calculations. For our measured $
(B-V)_{\rm J} = 0.99 \pm 0.03$ we obtain  
a mass of $0.8\,M_\odot$,
a luminosity of $\log{(L/L_\odot)} = -0.6$, 
and an effective temperature $\log{T_{\rm eff}}\,{\rm [K]} = 3.68$. 
According to \cite{Kenyon95.1} this temperature corresponds to a spectral
type of K2.5\,V.

\section{A preliminary orbit solution}\label{sect:orbit}

\begin{figure}
\begin{center}
\resizebox{8cm}{!}{\includegraphics{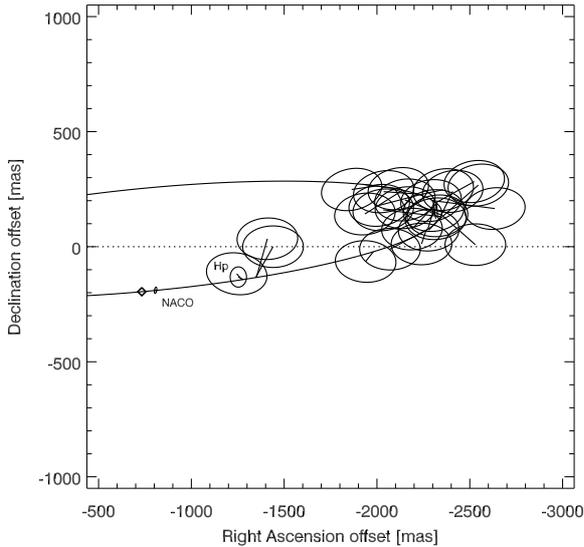}}
\caption{Apparent orbit of \gequ\,AB. 
The position of the primary is off to the left outside the displayed portion
of the orbit. 
The NACO measurement uncertainty ellipse is enlarged by a
factor of $2$ for clarity. 
The {\em Hipparcos} measurement is labeled `Hp'. 
A diamond marks the expected position of the companion at the date of the
{\em Chandra} observations.}
\label{fig:orbit}
\end{center}
\end{figure}

We combined our NACO measurement of \gequ 
with data from the WDS catalog\footnote{http://ad.usno.navy.mil/wds/data\_request.html} (WDS coordinates
$21103+1008$) and from {\em Hipparcos} 
to derive the preliminary orbit of \gequ\,AB shown in Fig.~\ref{fig:orbit}.
The measurements listed in the WDS cover the years between 1867 and 1994. 
Uncertainties on the WDS astrometry 
were estimated based on the scatter of separation 
($0.160^{\prime\prime}$) and position angle ($2.4^\circ$) 
relative to quadratic polynomial fits
of these quantities versus time.  The resulting reduced $\chi^2$ of the
orbital fit is $1.8$, and preliminary orbital elements with their formal
uncertainties are given in Table~\ref{tab:orbit}. Using the Hipparcos
parallax,  
%of $28.4 \pm 0.9$\,mas, 
we derive a total mass for 
the system \gequ~AB of $2.4 \pm 0.4\,M_\odot$. With the primary mass 
given from the HR diagram, 
the mass of component B is estimated to be $0.6 \pm 0.4\,M_\odot$
consistent with its photometry.

Given the large uncertainties of the astrometric data points, 
we do not put emphasis on the mass derived from the orbit.  
The orbit solution is, however, useful to 
predict the position of the secondary relative to the primary at the epoch 
of the X-ray observations, shown in Figs.~\ref{fig:images} and~\ref{fig:orbit}
and in Table~\ref{tab:separations}. 

\begin{table}
\begin{center}
\caption{Preliminary orbit for \gequ\,AB (see Fig.~\ref{fig:orbit}).}
\label{tab:orbit}
\begin{tabular}{lrr} \\ \hline
Element  & Unit       & Value \\ \hline
$a$      & [mas]      & $1600    \pm 100$ \\
$e$      &            & $0.56    \pm 0.05$ \\
$i$      & [$^\circ$] & $99.2    \pm 3.0$ \\
$\omega$ & [$^\circ$] & $196.2   \pm 4.0$ \\
$\Omega$ & [$^\circ$] & $95.0    \pm 3.0$ \\
$P$      & [yrs]      & $274.5   \pm 0.8$ \\
$T_{\rm 0}$ & [JD]       & $2475667  \pm 1000$ \\
\hline
\end{tabular}
\end{center}
\end{table}

\section{X-ray observations}\label{sect:xraydata}

$\gamma$\,Equ was observed under Obs-ID 9910
for $10$\,ksec with {\em Chandra} using the Advanced CCD Imaging Spectrometer 
(ACIS-I). 
The data analysis was performed with the CIAO software 
package\footnote{CIAO is made available by the CXC and can be downloaded 
from \\ http://cxc.harvard.edu/ciao/download/} version 4.2. 
We have applied standard filtering to the event file and applied the
good time interval file. 
We checked that there are no cosmic ray afterglows in the image around
the position of \gequ. 
Source detection was carried out with 
the {\sc wavdetect} algorithm \citep{Freeman02.1} using an image with 
spatial resolution of $0.25^{\prime\prime}$/pixel and a congruent, 
monochromatic exposure map for $1.5$\,keV.

\subsection{X-ray image}\label{subsect:xrayimage}

A zoom into the X-ray image for the 
$5^{\prime\prime} \times 5^{\prime\prime}$ around \gequ
is shown in Fig.~\ref{fig:images} (right). 
The proper motion corrected {\em Hipparcos} position of the roAp star is
marked with a large black x-point. % in Fig.~\ref{fig:images}. 
The only X-ray source detected in the vicinity of \gequ~A 
(marked with a small yellow circle in Fig.~\ref{fig:images}) 
is located $1.31^{\prime\prime}$ to the south-west ($\theta = 247.6^\circ$). 
The position angle is 
roughly coincident with the direction of the companion \gequ~B. 
Table~\ref{tab:separations} summarizes the separations between \gequ~A
and the companion measured with {\em Hipparcos}, NACO and {\em Chandra},
and compares them to the separation expected from the orbit solution.
\begin{table}
\begin{center}
\caption{Separation between \gequ~A and~B measured with different instruments at different epochs and 
corresponding predictions from orbit.}
\label{tab:separations}
\begin{tabular}{llcr}\\ \hline
Instrument & Date         & Measured          & Orbit solution \\ 
           & [yyyy/mm/dd] & [$^{\prime\prime}$] & [$^{\prime\prime}$] \\ \hline
Hipparcos  & 1991/04/02   & $1.26 \pm 0.04$   & $1.29$ \\
NACO       & 2007/06/15   & $0.826 \pm 0.003$ & $0.82$ \\
Chandra    & 2009/07/13   & $1.3 \pm 0.6^\dagger$     & $0.76$ \\   \hline
\multicolumn{4}{l}{$^\dagger$ This measurement relies on the identification of \gequ\,B with the} \\
\multicolumn{4}{l}{X-ray source; error estimate corresponds to formal accuracy of } \\
\multicolumn{4}{l}{ACIS-I astrometry.} \\
\end{tabular}
\end{center}
\end{table}
The expected position of the companion at the time of the {\em Chandra}
observation with respect to the proper motion corrected {\em Hipparcos} 
position of \gequ~A 
according to the orbit solution 
is marked with a small yellow x-point in Fig.~\ref{fig:images}. 
Given the difference in the separation between the expected position of \gequ~B
and the X-ray source it is appropriate to consider the astrometry. 

The positional accuracy of ACIS-I is $\sim 0.6^{\prime\prime}$ at
$90$\,\% confidence. 
We have performed source detection on the full {\em Chandra} image
(ACIS-0123) with the aim to better constrain the astrometry of the 
X-ray coordinates by cross-correlation with optical and infrared (IR) catalogs. 
However, only few of the $26$ X-ray sources detected across the ACIS-I 
image have known counterparts at other wavelengths, 
and all of them are either very weak ($<10$\,counts)
or at large off-axis angle. No systematic offset between optical/IR and
X-ray coordinates can be deduced. 
The adaptive optics image 
of the VLT does not allow for absolute astrometry and there are 
no other objects in the NACO field to which we could tie \gequ.

In the following we assume that \gequ\,B coincides with the X-ray source 
position. This seems a reasonable hypothesis, given the fact that
the accuracy of the absolute X-ray position may be uncertain by up to two 
pixels in Fig.~\ref{fig:images}, while the {\em Hipparcos} positional
accuracy ($\sim 1$\,mas) and proper motion accuracy ($1-2$\,mas/yr)
are superb. 
Shifting the image accordingly, 
the position of the primary, \gequ\,A, would then correspond to the 
large black circle in Fig.~\ref{fig:images}, 
using the separation and position angle 
from the orbit solution (Sect.~\ref{sect:orbit}). 

It seems that there is an 
excess of photons over the average background emission near the position 
of the primary, that we shall tentatively call `Source\,2'. 
This suggests that \gequ\,A is a faint X-ray source. 
However, this emission seems to be separated by more than the
expected $\sim 0.8^{\prime\prime}$ from the center of the brighter source 
associated with the secondary. 
The precise location of this emission
can not be determined given its small separation from the secondary's X-ray 
emission and the low photon statistics ($4$ counts; see below). 
As a test on the significance of `Source\,2' 
we computed the expected contribution from the wing of the point-spread-function 
of the X-ray source 
to the observed counts at the proper motion corrected 
{\em Hipparcos} position of \gequ\,A. 
In a $0.5^{\prime\prime}$ radius around this position $1.41$ counts are expected.
The Poisson probability of finding the observed $4$ photons in the same radius
is $5$\,\%.

\subsection{X-ray properties of the \gequ binary}\label{subsect:xrayspectrum}

When examining the properties of the X-ray source associated with the
companion we consider only photons within a $0.75^{\prime\prime}$ radius,
to avoid the (small) contribution from `Source\,2'. There are $89$ counts
in this photon extraction region.
No evidence for flaring is present. 

An individual response matrix and auxiliary response were extracted 
using standard CIAO tools. 
The background of ACIS is negligibly low.
We fitted the spectrum, rebinned to a minimum of $5$ counts per bin, 
in the XSPEC\,12.6.0 environment 
with a one-temperature thermal model subject to absorption ({\sc wabs $\cdot$ apec})
for a range of initial values for temperature and column density.
The elemental abundances relative to hydrogen are adopted from \cite{Wilms00.1} 
and are globally tied to $Z = 0.3\,Z_\odot$, 
such that the free parameters of
this model are temperature ($kT$), emission measure ($EM$) and column density 
($N_{\rm H}$).

The best fit parameters are given in Table~\ref{tab:xspec} together with 
uncertainties for a $68$\,\% confidence level. 
As expected for a nearby star without circumstellar matter, the best fit
has no absorption. The spectrum is formally also compatible with an absorbed
plasma of nearly one dex higher emission measure (see Table~\ref{tab:xspec}).
However, this solution goes along with quite low temperature and we consider it unlikely. 
The luminosity in the $0.3-8.0$\,keV band is $\log{L_{\rm x}}\,{\rm [erg/s]} = 28.06$,
where we have corrected for the flux in the PSF wings that were missed because of 
our limitation of the photon extraction radius. 

For the {\em ROSAT} FSC source ($0.020 \pm 0.008$\,cts/s) 
we derive with PIMMS\footnote{The Portable Interactive Multi-Mission Simulator (PIMMS) is accessible at http://asc.harvard.edu/toolkit/pimms.jsp} 
a similar X-ray luminosity of $\log{L_{\rm x}}\,{\rm [erg/s]} = 28.3$.
For this estimate that refers to the $0.12-2.48$\,keV band 
we have assumed a Raymond-Smith 
spectrum with the temperature derived from the {\em Chandra} data. 
\begin{table}[t]
\begin{center}
\caption{XSPEC fit results for the X-ray source associated with \gequ~B.
The free parameters of the one-temperature APEC model are column density ($N_{\rm H}$),
temperature ($kT$) and emission measure ($EM$). 
Uncertainties are $68$\,\% confidence levels.}
\label{tab:xspec}
\begin{tabular}{rrrr}\hline
$\chi^2_{\rm red}$ (dof) & $\log{N_{\rm H}}$     & $kT$   & $\log{EM}$        \\ 
                         & [${\rm 10^{22}\,cm^{-2}}$]     & [keV]  & [${\rm cm}^{-3}$] \\\hline
0.76 ( 12)               &  $0.0^{+0.44}_{-0.0}$ &   $0.52^{+0.62}_{-0.27}$ & $51.2^{+52.0}_{-51.0}$   \\ \hline
\end{tabular}
\end{center}
\end{table}
 
The total of only four photons ascribed to `Source\,2' prohibits any firm
conclusions on its X-ray properties. 
The photon energies are $0.7, 0.9, 1.2$ and $1.8$\,keV, 
%% PHOTON ENERGIES FROM ACIS_EVT2_RED_1.0PIX.FITS
suggesting spectral hardness similar to that of 
the emission associated with \gequ\,B.  
From the arrival time distribution of the photons we do 
not infer evidence for flaring. % (see Fig.~\ref{fig:time_energy}). 
Assuming the same X-ray temperature as for 
\gequ\,B we obtain with PIMMS a luminosity of $\log{L_{\rm x}}\,{\rm [erg/s]} = 26.6$.

\section{Discussion and conclusions}\label{sect:discussion}

The atmospheres of the roAp stars can be termed pathological.
\cite{Cowley04.1} and \cite{Fivet07.1} called 
attention to the possible presence of unstable elements (Tc and Pm) in
some roAp stars. 
The longest-lived Pm isotope has a half-life of only 17.7 years.  
Accelerated particles may lead to nucleosynthesis of short-lived elements
on the surface of Ap stars \citep{Goriely07.1}. 
It is a well-known fact that charged particles are accelerated 
during stellar and solar flares \citep{Hudson95.1}. 
These particles thermalize their energy, giving rise to
high-energy X-ray and $\gamma$-ray emission. While the latter has remained undetectable
on stars other than the Sun, part if not all of the X-ray emission from magnetically active
stars is usually attributed to such flare events. 
Therefore, it is not implausible to attribute the presence of isotopes with short half-lifetime 
in roAp stars to flare activity. The detection or failure to find X-rays from these
stars may, therefore, provide critical information on the state of the 
plasma in their atmospheres. 

Our search of several X-ray catalogues for X-ray sources coinciding with 
a roAp star, brought forth only one object, \gequ. 
We have now shown in a dedicated {\em Chandra} observation that the bulk of 
this X-ray emission must be ascribed to a late-type
companion. The photometry of this object points to an early-K spectral type, 
assuming coevality with \gequ\,A. The physical pairing
with the primary is corroborated by clear evidence for orbital motion. 
The X-ray luminosity (${\log{L_{\rm x}} = 28.1}$), X-ray temperature 
($kT = 0.5$\,keV) and fractional X-ray luminosity
[$\log{(L_{\rm x}/L_{\rm bol}) = -5}$] are 
in accordance with the expectation for the age and spectral type of \gequ\,B;
cf. \cite{Preibisch05.3, Schmitt90.1}. 

More important is the possible identification of faint X-ray emission from 
the roAp star. The enhanced number of photons to the north-east of the secondary
has low probability to be a statistical fluctuation but its association with
\gequ\,A could not be established with certainty due to uncertain astrometry. 
If this faint emission is attributed to \gequ\,A, its fractional X-ray luminosity
is extremely low, $\log{(L_{\rm x}/L_{\rm bol}) = -7.9}$. 
 
The mass and age of \gequ\,A are similar to the normal, i.e. non-peculiar,
A-type star Altair shown by \cite{Robrade09.1} to be an X-ray emitter with
$\log{(L_{\rm x}/L_{\rm bol})} = -7.4$. Contrary to \gequ\,A, 
Altair is a fast rotator and its X-ray properties are ascribed to compact structures
at low latitudes, i.e. its X-ray emission can reasonably be explained by a weak
solar-like corona. The extremely long period derived from the magnetic field variations 
for \gequ\,A makes such an interpretation difficult as it translates to extreme
values for the Rossby number ($R_0 = \tau_{\rm conv}/P_{\rm rot}$) that have never
been investigated. 
Extrapolation of the empirical relation between X-ray emission and rotation rate 
for MS stars of masses $<1.3\,M_\odot$ \citep{Pizzolato03.1},
if valid for stars of the mass of \gequ\,A,  
would predict several orders of magnitude fainter X-ray luminosity than inferred from 
the detection of $4$ counts in our {\em Chandra} observation. 
Therefore, for \gequ\,A, either the magnetic period 
is not identical with the rotation period or the X-ray emission mechanism is 
not due to a solar-like dynamo on the roAp star. 

While residual dynamo activity connected
to a (shallow) convection zone is the only viable mechanism for intrinsic
X-ray production in normal intermediate-mass stars,  
magnetically confined wind-shocks (MCWS) provide an alternative in the
case of Ap stars. MCWS have been invoked as explanation
for the X-ray emission from the Ap star IQ\,Aur \citep{Babel97.1}. 
The X-ray luminosity of IQ\,Aur ($\log{L_{\rm x}} = 29.6$) is about 
$1000$ times higher than that of \gequ\,A 
(see Sect.~\ref{subsect:xrayspectrum}). According to \cite{Babel97.1} for the
MCWS model $L_{\rm x}$ scales with the magnetic field strength ($B_*$), 
the mass loss rate ($\dot{M}_{\rm W}$) and the terminal wind velocity 
($v_\infty$) as
\begin{equation}
L_{\rm x} \simeq 2.6 \cdot 10^{30} {\rm erg/s} \cdot (\frac{B_*}{1\,{\rm kG}})^{0.4} \xi
\end{equation}
and
\begin{equation}
\xi = (\frac{\dot{M}_{\rm W}}{10^{-10}\,{M_\odot{\rm /yr}}})^\delta (\frac{v_\infty}{10^3\,{\rm km/s}})^\epsilon
\end{equation}
with $\delta = 1$ and $\epsilon \approx 1...1.3$. In the case of \gequ\,A, 
the value measured for the maximum longitudinal field is 
$B_{\rm l, max} \approx 1$\,kG 
\citep{Bychkov06.1}. According to \cite{Auriere07.1}, the surface field
of a dipole is $B_{\rm s} \geq 3.3 B_{\rm l, max}$. We further assume 
$v_\infty \simeq 600$\,km/s, the value derived for IQ\,Aur by 
\cite{Babel97.1}. With these numbers 
the mass loss rate required to sustain the X-ray luminosity estimated from our 
observation is $2 \cdot 10^{-14}\,M_\odot{\rm /yr}$.  
\cite{Michaud86.1} showed that the CP phenomenon appears
only for $\dot{M}_{\rm W} < 10^{-12}\,M_\odot{\rm /yr}$ and that 
mass loss rates
on magnetic Ap stars may, indeed, be as small as $10^{-15}\,M_\odot{\rm /yr}$. 
If we further set for the velocity of the shock ($v_{\rm sh}$) 
forming from the collision of the winds from the two hemispheres in the 
equatorial plane $v_{\rm sh} \simeq v_\infty$, the strong shock condition 
implies a shock temperature of $\approx 0.35$\,keV. As argued in 
Sect.~\ref{subsect:xrayspectrum} 
the energies of the four photons from `Source 2'
hint at a similar spectral temperature as measured for \gequ~B, $0.52$\,keV,
which is roughly consistent with the above estimate. 
Clearly, these order of magnitude estimates that rely on a reasonable guess
and approximations 
for some parameters can serve as an argument for plausibility of
the MCWS scenario on \gequ\,A but do not demonstrate its actual presence
on this star. 

An alternative explanation for the X-ray photons that we ascribed to \gequ\,A 
would be an additional unseen companion. Such an object would also solve the problem of the
discrepancy between the separations of \gequ\,AB measured in the near-IR from the
NACO image on the one hand and in the X-ray image (\gequ~B vs. `Source\,2') on the
other hand; see Fig.~\ref{fig:images}.  
This close companion would have to have a separation of only $\sim 0.4^{\prime\prime}$
from \gequ\,A.
At this separation $90$\,\% completeness is reached in the NACO image 
for companions with a flux of at
least $1$\,\% of the primary (Fig.~4 of \cite{Schoeller10.1}), i.e.   
any such hidden companion in our case would likely be fainter than $K \sim 9$\,mag 
because it escaped detection with NACO. 

Clearly, the possible detection of X-ray emission from the roAp star \gequ\,A
based on $4$ photons needs to be corroborated by future deeper observations. 
If confirmed, the X-ray emission may be a diagnostic for the energetic events
that are believed to be at the origin of the production of unstable elements on 
some roAp stars. 
The fact that radioactive elements are not currently seen on \gequ\,A itself 
does not exclude this scenario. 
The exotic elements may be concentrated in localized regions on the stellar 
surface which, as a consequence of its very slow rotation, may remain  
hidden to the observer for decades. Spots of rare-earth elements are well-established
in the Ap star HR\,465 \citep[e.g.][]{Cowley81.1, Rice88.1}, 
for which the presence of Pm\,II has been suggested.
HR\,465 shares with \gequ\,A the characteristic of a very long rotation period 
($\sim 22-24$\,yrs) but it is not a roAp star.

\begin{acknowledgements}
BS wishes to thank Loredana Prisinzano and Ettore Flaccomio for stimulating 
discussions. 
We would like to thank Serge Correia for contribution to the NACO data reduction. 
We appreciate the valuable feedback of the referee, M.Auri\`ere. 
BS acknowledges financial support by ASI-INAF agreement I/009/10/0. 
We have used data and the software CIAO 
obtained from and provided by the {\em Chandra} X-ray Observatory Center, 
which is operated by the Smithsonian Astrophysical Observatory for and on behalf 
of the National Aeronautics Space Administration under contract NAS8-03060.
Observations were also made at the ESO Paranal Observatory with the VLT under 
programme ID 079.D-0537(A). 
Moreover this research has made use of the 
{\em Bag of Stellar Tracks and Isochrones} (BaSTI) database 
and of the Washington Double Star Catalog maintained
at the U.S. Naval Observatory.
\end{acknowledgements}

\bibliographystyle{aa} %aa.bst
\bibliography{16265,mnemonic}

\end{document}